# Towards quantum dot arrays of entangled photon emitters

Gediminas Juska[*,1], Valeria Dimastrodonato[1], Lorenzo O. Mereni[1], Agnieszka Gocalinska[1] and Emanuele Pelucchi[1]

[1]Tyndall National Institute, University College Cork, Lee Maltings, Cork, Ireland

[*]Corresponding author: gediminas.juska@tyndall.ie

**To make photonic quantum information a reality[1,2], a number of extraordinary challenges need to be overcome. One of the outstanding challenges is the achievement of *large arrays* of reproducible "entangled" photon generators, maintaining the compatibility with integration with optical devices and detectors[3,4,5]. Semiconductor quantum dots (QDs) are potentially ideal for this. They allow generating photons on demand[6,7] without relying on probabilistic processes[8,9]. Nevertheless, most QD systems are limited by the intrinsic lack of symmetry, which allows to obtain only a small number (typically 1/100 or worse) of good dots per chip. The recent retraction of Mohan et al.[10] seemed to question the very possibility of matching site-control and high symmetry. Here we show that with a new family of (111) grown pyramidal site-controlled $InGaAs_{1-\delta}N_\delta$ QDs, it is possible to overcome previous difficulties and obtain areas containing as much as 15% of polarization-entangled photon emitters, showing fidelities as high as 0.721±0.043.**

The idea underlining the principle of entangled photon emission with QDs relies on fundamental quantum physics: particle indistinguishability generates a superposition state when two energetically nearly degenerate quantum levels are populated at the same time. In QDs, entanglement resides in polarization of two photons emitted during the cascaded biexciton-exciton recombination[11]. Here one difficulty arises: when the two excitonic levels are not perfectly degenerate (i.e. there is a fine structure splitting, FSS), the entanglement in the emission persists, but a phase term between the two (linearly) polarized photons proportional to both energy and time is introduced. This results in a relative rotation of the two photon polarizations (not constant in time) making entanglement substantially impossible to be detected in a simple way[12].

All currently reported QD systems allowing entangled photon emission tend to present a large FSS, fundamentally allowing only a few (post-growth selected) QDs on a semiconductor wafer as good sources, while till now no entangled photon emission has been demonstrated with any system allowing accurate QD position control. By all means it is text book physics, that, to observe level degeneracy, one needs a symmetric confinement. As discussed in a number of contributions, growth along the [111]B crystallographic direction shows ideally a $C_{3V}$ symmetry[13,14,15], which should allow the realisation of large arrays of position-controlled entangled photon emitters. In practise, however, a relatively broad range of FSS can be found on existing (111) systems, as for example in our pyramidal QDs. An efficient way to overcome asymmetry related issues was obtained by exposing a QD layer to unsymmetrical dimethylhydrazine (U-DMHy, a standard source of nitrogen during metalorganic vapour phase epitaxy, MOVPE) during the QD formation process[16,17]. It was consistently observed that the presence of U-DMHy within a certain range of growth conditions helped to improve the symmetry of the QDs, enabling reproducible fabrication of nanostructures with a FSS consistently below our detection limit of ~4 µeV.

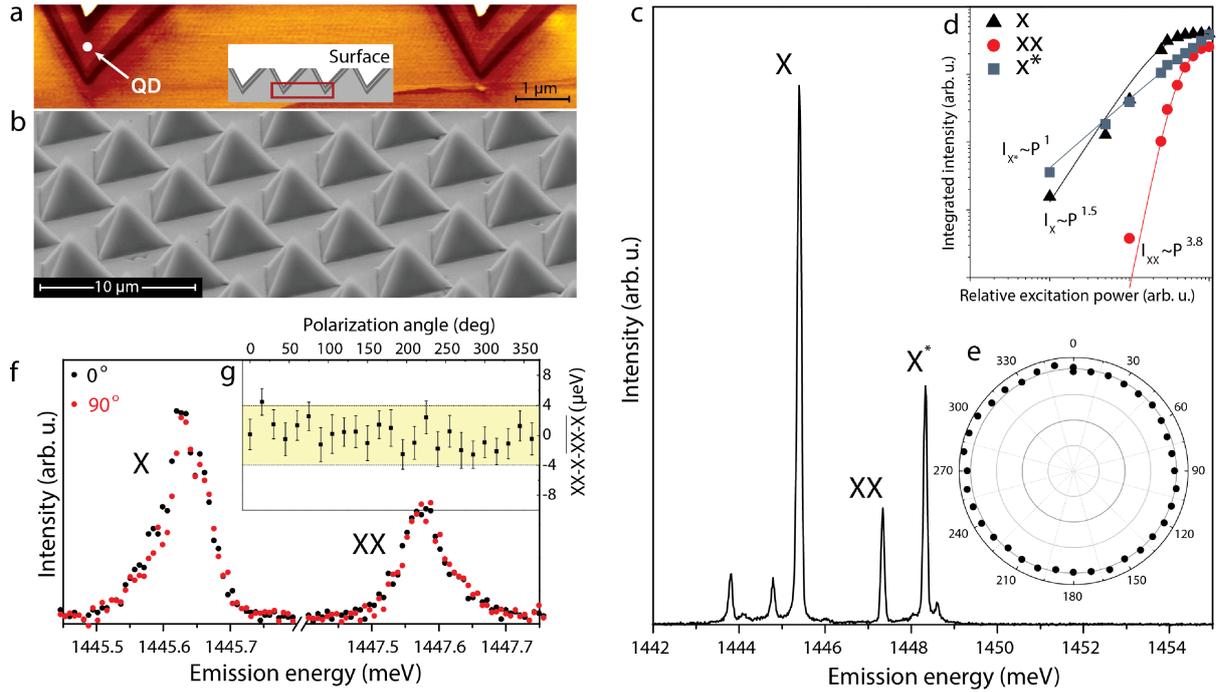

**Figure 1. Pyramidal site-controlled QDs of a high symmetry. a**, AFM image of a cleaved QD sample in a side-view. The contrast is provided by the different oxidation rates of pure GaAs and Al containing layers. The rectangle in a full cross-section view sketch indicates the position depicted by the AFM image. **b**, The tilted HRSEM image of the sample after the post-growth substrate removal procedure. **c**, The typical photoluminescence spectrum of a QD. Exciton, biexciton and generic charged exciton transitions are identified as X, XX and X$^*$, respectively. **d**, Example of excitation power dependency. **e**, Integrated intensity of exciton transition as a function of polarization angle showing isotropic in-plane distribution of photoluminescence polarization. **f**, Representative exciton and biexciton spectra taken at perpendicular linear polarization angles. **g**, A more precise FSS measurement procedure. Error bars show the standard deviation in the specific measurement, and 4 µeV bars represent the typical FSS detection range.

The investigated $In_{0.25}Ga_{0.75}As_{1-\delta}N_\delta$ QDs were grown by MOVPE in 7.5 µm pitch tetrahedral recesses etched in (111)B oriented GaAs. An atomic-force microscopy image (Fig. 1a) of a cleaved sample in side-view shows the epitaxial layer structure (see "methods") where a QD is located at the central axis of the recess, within GaAs barriers. Photoluminescence extraction enhancement was achieved by selectively removing (back-etching) the substrate (Fig. 1b). The typical Lorentzian linewidth of the exciton transition was found to be 80±15 µeV in our first samples. As we discuss in supplementary material, this broadening does not represent a fundamental limitation.

Figure 1c shows a typical photoluminescence spectrum of single QDs investigated in this work. The characteristic feature is a biexciton transition at higher energy (antibinding biexciton) in entangled photon emitters, always accompanied by the presence of a charged exciton at higher energy. The significance of the spectrum must be stressed, as it acts as a very precise and quick indicator preselecting QDs that emit polarization-entangled photons (see supplementary material for more comments).

Figure 1e shows an isotropic linear polarization distribution proving QDs as sources of unpolarized light. Photoluminescence intensity dependence on excitation power is depicted in Fig. 1d. It acts as a preliminary indicator of the excitonic transitions. Despite the deviation from the ideal linear and quadratic power dependence, the type of transitions was unambiguously confirmed by photon correlation measurements. Average lifetime values are consistent with the attributed transition types, as well: 1.8±0.6 ns and 0.9±0.15 ns for exciton and biexciton, respectively. Figure 1f illustrates exciton and biexciton transitions measured/filtered at perpendicular linear polarization angles. No visual difference can be identified between the spectra. In the presence of low symmetry (i.e with a FSS), both peaks should be composed of two energetically distinguishable linearly polarized components (typically referred to as *H* and *V*). Here no particular crystallographic direction can consistently be associated with *H* and *V* components, as the origins of the FSS are related to random effects, and not to a shape elongation along certain directions as in self-assembled QDs. *H* and *V* in our work only indicate 0 and 90 deg angles with respect to our linear polarizer. To obtain the exact value of the FSS we apply a well-known FSS measuring procedure which involves taking a set of polarized spectra at smaller polarization angle steps (Fig. 1g)[18].

If the intermediate exciton level is degenerate, during the recombination cascade, the emitted pair has a polarization entanglement expressed by the Bell state $|\psi\rangle = \frac{1}{\sqrt{2}}(|L_{XX}R_X\rangle + |R_{XX}L_X\rangle)$. Using Jones vectors, the state can be rewritten as $|\psi\rangle = \frac{1}{\sqrt{2}}(|H_{XX}H_X\rangle + |V_{XX}V_X\rangle) = \frac{1}{\sqrt{2}}(|D_{XX}D_X\rangle + |A_{XX}A_X\rangle)$, with *L(R)* – circularly left (right) hand, *H(V)* – horizontally (vertically), *D(A)* – diagonally (antidiagonally) polarized photons, *XX* – biexciton, *X* – exciton. As a consequence, an ideal source of polarization entangled photons should display a perfect correlation measured in linear and diagonal bases and a perfect anticorrelation in a circular basis. Figure 2a presents polarization-resolved second-order correlation functions taken in the before-mentioned polarization bases on a representative dot. Clear bunching is observed in co-polarized linear and diagonal and counter-polarized circular photon correlation curves (these non-classical correlations can be lost as soon as the FSS is measurable (e.g. ~3µeV – see supplementary material)). The degree of correlation of unpolarized source can be obtained directly from the experimental data by $C_{basis} = \left(g^{(2)}_{xx,x} - g^{(2)}_{xx,\bar{x}}\right) / \left(g^{(2)}_{xx,x} + g^{(2)}_{xx,\bar{x}}\right)$, with $xx(x)$ - polarization of biexciton (exciton) and $\bar{x}$ - orthogonal polarization of exciton[19]. Figure 2b plots degrees of correlations at different excitation delays for the same QD. This can be used to calculate the fidelity of the entangled state $|\psi\rangle = \frac{1}{\sqrt{2}} \left(|H_{XX} H_X\rangle + |V_{XX} V_X\rangle\right)$ (see "methods"). In the below described statistical distribution of QDs, we use the fidelity as a figure of merit to evaluate entanglement, as the experimental procedure is significantly less time-consuming than reconstructing a full density matrix. This is accepted by the scientific community as a proof of entanglement[20,21] (indeed the method gives an equivalent result to a full quantum state tomography). The calculated fidelity value of 0.670±0.035 in the given example (Fig. 2a(2b)) exceeds the maximum limit of classically correlated light (0.5) by nearly five standard deviations indicating the entangled nature of the emitted photons. The highest obtained value over the sample was 0.721±0.043. As discussed elsewhere[19], in general, a variety of effects, such as spin scattering, background light, and dephasing, are responsible for reduced fidelities.

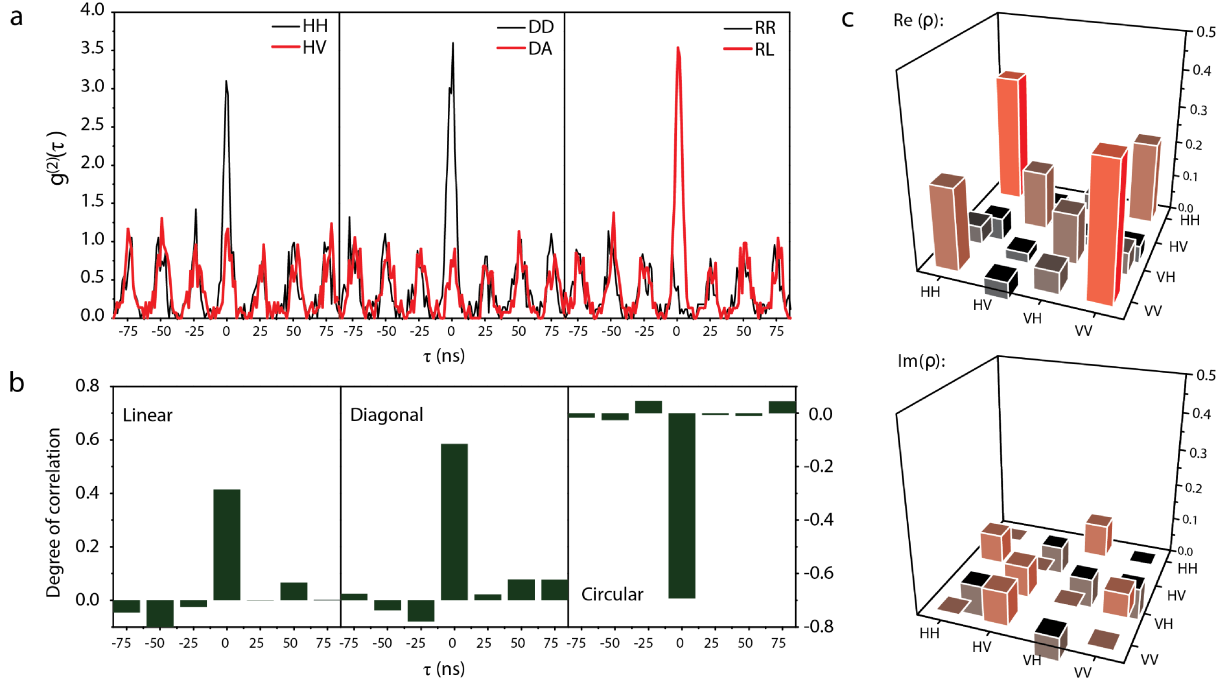

**Figure 2. Polarization-entangled photons. a**, Representative second-order correlation functions taken for co-polarized and cross-polarized biexciton-exciton photons in three different polarization bases for a specific dot as discussed in the text. Strong bunching of co-polarized linear and diagonal and cross-polarized circular $g^{(2)}$ functions clearly identifies non-classical correlations. As a rule of thumb, when curves are normalized to the respect of side peaks, entanglement can be attested by analysing each pair of $g^{(2)}$ curves – the area under each bunching peak (around zero delay) has to be more than twice the area under the corresponding peak (ideally antibunched) from the other curve – $g^{(2)}_{bunched} > 2 g^{(2)}_{antibunched}$ (see "methods"). **b**, Degrees of correlation in different polarization bases for the same measurement/dot. **c**, Representative density matrix reconstructed by quantum state tomography procedure (note that a different QD was used in these measurements from that used in the measurements for (a) and (b)).

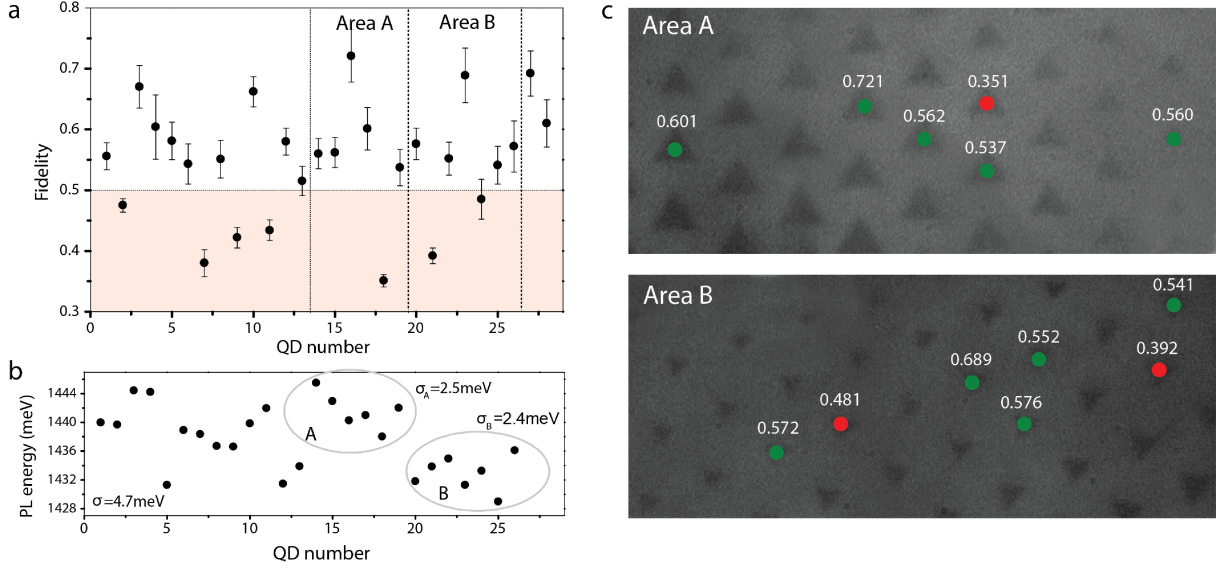

**Figure 3. High density of QDs emitting polarization entangled photons. a**, The distribution of fidelity values of all measured QDs. The limit of 0.5 for classically correlated light is passed in the majority cases. **b**, The distribution of emission energy of the respective QDs in (a) with standard deviation of 4.7 meV. Locally (e.g. areas A and B) it can be reduced to ~2.5 meV. **c**, Images of two randomly selected areas (A and B in the previous graphs) taken with the sample surface visualization system installed in the set-up. The green and red spots identify all pre-selected QDs for polarization-entanglement characterization. The green spots identify QDs with fidelity of entangled state (values are above the spots) >0.5, while red ones <0.5.

In general, a two photon polarization state is fully described by a density matrix. Figure 2c plots real and imaginary parts of the density matrix reconstructed from the experimental data (of the representative QD analysed) using a quantum state tomography procedure[22]. The density matrix is represented using $|HH\rangle$, $|HV\rangle$, $|VH\rangle$ and $|VV\rangle$ as a basis set. The presence of non-zero outer elements $|HH\rangle\langle VV|$ and $|VV\rangle\langle HH|$ in the minor diagonal exceeding $|HV\rangle\langle HV|$ and $|VH\rangle\langle VH|$ is due to the superposition of photon pair wavefunction – a signature of entanglement. In the absence of superposition, only the outer diagonal elements would be present, indicating classically correlated light (provided that the orientation of linearly polarized component due to a FSS is coincident with the corresponding linear polarization projection). In addition to the fidelity $F$, entanglement can be validated by passing a number of other tests such as concurrence $C$[23], tangle $T=C^2$ [24] and the Peres criterion[25]. They all confirm in this (and the other) cases non classical light: $F=0.582\pm0.031$ (>0.5), $C=0.158\pm0.020$ (>0), $T=0.025\pm0.020$ (>0), Peres criterion $-0.119\pm0.035$ (<0).

To demonstrate the high density of entangled photon emitters, a number of randomly chosen QDs, however preselected mostly by the photoluminescence spectrum (Fig. 1c) (as discussed above), was characterized in terms of fidelity. Figure 3a presents the measured distribution of fidelity values indicating that 75% of the preselected QDs passed the limit of 0.5 for classically correlated light. Figure 3c consists of the two images obtained by the sample surface imaging system of two randomly selected areas (noted as A and B). Green spots indicate QDs that have fidelity of the entangled state >0.5, while the red ones ≤0.5. In the both cases, the percentage of QDs with fidelity >0.5 is at least 15% of the overall QD field. Such high concentration of QDs emitting entangled photons corresponds mostly to those sample areas where the substrate was not fully removed during back-etching. The density of QDs emitting polarization-entangled photons drops in the central areas (where the full substrate removal procedure was more effective), probably due to the prolonged etching which might have degraded the QD quality. This assumption is supported by the fact that better statistics of entangled photon emitters' distribution in the central area was achieved in a successful reproduction of the sample.

Finally, another remarkable feature of our site-controlled quantum dot family also worth mentioning: the emission energy uniformity. Figure 3b shows emission energy distribution of the dots selected for the fidelity measurements. The total area of interest was 4x3 mm$^2$. Although the total standard deviation is 4.7 meV, over a short spatial range its distribution is reduced to ±2.5 meV (areas A and B).

Concluding, the relevance of our findings is the high density of QDs that emit polarization-entangled photons without external manipulation of the electronic states. Areas containing at least 15% of entangled photon emitters could be easily found. It is a major improvement comparing to other types of epitaxial QDs where only a small number of all QDs can be direct emitters of entangled photons. This opens already the possibility to build large arrays of entangled photon emitting units that could be integrated into future quantum computation devices. Moreover, the control of pyramidal site-controlled QDs could be in the future enhanced by local metallic gates that could be used to inject carriers and/or electrostatically manipulate excitonic states. The demonstrated apex-up geometry and micron pitched pyramids are actually beneficial for such design. In addition, local strain manipulation[26] could also be utilised. All together these techniques (and further epitaxial optimisation) should allow the development of "perfect" arrays of entangled photon emitters, where all devices would act as "good" emitters, effectively contributing to the construction of "real world" quantum computation with flying qubits.

## Methods

### QD growth

Quantum dots were grown by MOVPE in 7.5 μm pitch tetrahedron recesses etched by wet chemical etching procedure in (111)B oriented GaAs substrate. Standard MOVPE precursors, namely trimethyl(-gallium, -indium, -aluminium), unsymmetrical dimethylhydrazine (U-DMHy) and arsine (AsH$_3$), were used, in a reactor where special care is taken with impurity reduction[27]. A sequence of epitaxial layers was grown: a buffer layer of GaAs followed by Al$_x$Ga$_{1-x}$As alloy of gradually increasing Al from 0.3 to 0.75. Then an etching-resistant layer of Al$_{0.75}$Ga$_{0.25}$As was grown in order to enable selective post-growth substrate removal procedure. A bottom cladding layer of Al$_{0.55}$Ga$_{0.45}$As was topped with confining GaAs barrier of 100 nm thickness. The In$_{0.25}$Ga$_{0.75}$As$_{1-\delta}$N$_\delta$ QD layer of 0.85 nm nominal thickness was grown at 730°C temperature (thermocouple reading). Due to the interplay of anisotropic growth and capillarity effects[28,29], a QD shape is determined by the underlying GaAs self-limiting profile. An estimated, supported by theoretical predictions, diameter of the QD base is ~70 nm. The $In_{0.25}Ga_{0.75}As_{1-\delta}N_\delta$ layer is topped with 70 nm thick GaAs and $Al_{0.55}Ga_{0.45}As$ barriers.

U-DMHy presumably acts as a surfactant layer, effectively reducing segregation related growth disorder. Moreover, the QD structure is not grown immediately after a buffer deposition as done elsewhere[10], but after a number of "thick" cladding layers have developed: this ensures that stationary lateral confinement, induced by the self-limited profile, is reached and structural and compositional dot symmetry fully exploited. It should be noted that we thoroughly investigated the time/epi-thickness evolution with growth time, and experimentally observed that growth transients are unexpectedly slow (as much as ~hundred nm, before equilibrium is reached).

**Substrate removal**

A relevant issue of as-grown samples is the photoluminescence extraction efficiency which leads to unfavourable experimental conditions. To overcome these problems, the substrate was selectively etched away[30] leaving an array of apex-up pyramids standing on a supporting substrate. The sample was attached by thermocompression gold bonding to a (100)GaAs sample before the "all chemical" substrate removal procedure was performed. The photoluminescence extraction enhancement is due to the pyramidal shape which acts as a lens and, in a minor way, is also due to the reflecting gold layer under the base of pyramid. In major cases, it is an essential procedure to get a significant PL signal of QDs.

**Measurement and setup details**

Photoluminescence data was taken in the conventional micro-photoluminescence set-up, which enables access to individual QDs. The sample was cooled down to 7 K by a closed-cycle, low vibrations microPL helium cryostat (ARS cryo). QDs were excited non-resonantly with semiconductor laser diode emitting at 635 nm with a repetition rate of 40 MHz. Backscattered light was collected through a 50x objective with numerical aperture of 0.55. Exciton and biexciton transitions were filtered for correlation measurements by two monochromators using spectral resolution of 0.5 meV and 0.2 meV, respectively. Each filtered transition was divided by a polarizing beamsplitter and sent to silicon avalanche photodiodes (APD). Typical number of counts per second during the measurements of entangled photons were 3000 and 700 for exciton and biexciton transitions, respectively, yielding to the detection of ~810 entangled photon pairs per hour. Four synchronized sequences of APD signals were fed to photon counting module and analysed in order to build four correlation curves. Such a four detector set-up allowed measuring a complete polarization basis simultaneously, increasing precision of the quantum state tomography procedure by reduced statistical errors (e.g., present due to the sample drifting, excitation intensity fluctuations), which are harder to avoid during a long experimental procedure if only information of a single polarization basis state intensity is collected. Intensity of each of the states directly involved in the tomographic analysis (16 in total) was obtained by converting measured second order correlation function values to probabilities. Polarization bases were chosen by using an appropriate combination of half- or quarter- waveplates and polarizers.

The FSS was measured by placing a polarizer in front of the spectrometer entrance and rotating half-waveplate by fixed steps. The corresponding exciton and biexciton transitions were fitted by Lorentzian fits and peak positions plotted as functions of polarization angle. In the presence of a small FSS, the curves typically follow sinusoid function from which FSS value can be obtained directly. Such fitting procedure combined with the 18 μeV experimental resolution improves our total measurement resolution to at least 4 μeV.

The second order correlation function value was obtained by dividing all events at zero delay peak (from -12.5 ns to 12.5 ns) by the mean value of events of the side peaks. For the calculations, raw data was used without background subtraction. The fidelity $F$ of the entangled state $|\psi\rangle = \frac{1}{\sqrt{2}}(|H_{XX}H_X\rangle + |V_{XX}V_X\rangle)$ was calculated from the degrees of correlations ($C_{basis}$) obtained in linear ($L$), diagonal ($D$) and circular ($C$) polarization basis[27]: $F = (1 + C_L + C_D - C_C)/4$, where $C_{basis} = (g^{(2)}_{xx,x} - g^{(2)}_{xx,\bar{x}})/(g^{(2)}_{xx,x} + g^{(2)}_{xx,\bar{x}})$. Entanglement is attested when $F > 0.5$ (if $|C_L| \approx |C_D| \approx |C_C|$, as a rule of thumb, $g^{(2)}_{xx,x} > 2g^{(2)}_{xx,\bar{x}}$ for the bases where bunching is expected between co-polarized photons).

## Acknowledgements

This research was partly enabled by the Irish Higher Education Authority Program for Research in Third Level Institutions (2007-2011) via the INSPIRE programme, by Science Foundation Ireland under grants 05/IN.1/I25, 10/IN.1/I3000 and 08/RFP/MTR/1659, and EU FP7 under the Marie Curie Reintegration Grant PERG07-GA-2010-268300. We are grateful to K. Thomas for his support with the MOVPE system and RJ Young for his essential help in setting up the correlation set-up in the early stages of this project.

## Author contributions

GJ carried out optical characterisation of the samples and data analysis, wrote the manuscript with EP. LOM assisted in optical characterisation and data analysis. VD and AG participated in the production of the samples, processing and microscopy characterization. EP conceived the study, and participated in its design and coordination, writing the manuscript. All authors commented on the final manuscript.

## Additional information

**Competing Financial Interests:** The authors declare no competing financial interests. Correspondence and requests for materials should be addressed to GJ.